%% file: ifacconf.tex
\documentclass{ifacconf}
\usepackage{graphicx}      % include this line if your document contains figures
\usepackage{natbib}        % required for bibliography
\usepackage{tikz}
\usetikzlibrary{positioning,arrows,calc,math,angles,quotes,shadings}
\usepackage{amsmath}
\usepackage{amssymb}
\usepackage{subfloat}
\makeatletter
\def\endfigure{\end@float}
\def\endtable{\end@float}
\makeatother

\let\ifacconfcaptionwidth\captionwidth
\usepackage[caption=false]{subfig}
\let\captionwidth\ifacconfcaptionwidth
\usetikzlibrary{tikzmark, positioning, fit, shapes.misc}
\usepackage{xurl}
\usetikzlibrary{decorations.pathreplacing, calc}

\DeclareMathOperator*{\argmin}{arg\,min}
\DeclareMathAlphabet{\mathpzc}{T1}{pzc}{m}{it}
\tikzset{brace/.style={decorate, decoration={brace}},
  brace mirrored/.style={decorate, decoration={brace,mirror}},
}
\begin{document}
\begin{frontmatter}

\title{Economic Predictive Control with Periodic Horizon for Water Distribution Networks } 

\thanks[footnoteinfo]{This work is funded by Independent Research Fund Denmark (DFF). We acknowledge Verdo company, Peter Nordahn, and Steffen Schmidt for providing us with the EPANET model and the network information.}

\author[First]{Mirhan Ürkmez} 
\author[First]{Carsten Kallesøe}
\author[First]{Jan Dimon Bendtsen}
\author[First]{John Leth}

\address[First]{\textit{Aalborg University, Fredrik Bajers Vej 7c, DK-9220 Aalborg,
Denmark}\\ \textit{
(e-mail: \{mu,csk,dimon,jjl\}@es.aau.dk)}}

\input{abs}


\end{frontmatter}
%===============================================================================

\input{intro}
\input{system}

\input{Control}

\input{exp}
\input{conc}

\bibliography{ifacconf}             % bib file to produce the bibliography
                                                     % with bibtex (preferred)
                                                   
%\begin{thebibliography}{xx}  % you can also add the bibliography by hand

%\bibitem[Able(1956)]{Abl:56}
%B.C. Able.
%\newblock Nucleic acid content of microscope.
%\newblock \emph{Nature}, 135:\penalty0 7--9, 1956.

%\bibitem[Able et~al.(1954)Able, Tagg, and Rush]{AbTaRu:54}
%B.C. Able, R.A. Tagg, and M.~Rush.
%\newblock Enzyme-catalyzed cellular transanimations.
%\newblock In A.F. Round, editor, \emph{Advances in Enzymology}, volume~2, pages
%  125--247. Academic Press, New York, 3rd edition, 1954.

%\bibitem[Keohane(1958)]{Keo:58}
%\newblock \emph{Power and Interdependence: World Politics in Transitions}.
%\newblock Little, Brown \& Co., Boston, 1958.

%\bibitem[Powers(1985)]{Pow:85}
%\newblock Is there a way out?
%\newblock \emph{Harpers}, pages 35--47, June 1985.

%\bibitem[Soukhanov(1992)]{Heritage:92}
%\newblock \emph{{The American Heritage. Dictionary of the American Language}}.
%\newblock Houghton Mifflin Company, 1992.

%\end{thebibliography}

\end{document}

%% file: abs.tex
\begin{abstract}
This paper deals with the control of pumps in large-scale water distribution networks with the aim of minimizing economic costs while satisfying operational constraints. Finding a control algorithm in combination with a model that can be applied in real-time is a challenging problem due to the nonlinearities presented by the pipes and the network sizes. We propose a predictive control algorithm with a periodic horizon. The method provides a way for the economic operation of large water networks with a small linear model. Economic Predictive control with a periodic horizon and a terminal state constraint is constructed to keep the state trajectories close to an optimal periodic trajectory. Barrier terms are also included in the cost function to prevent constraint violations. The proposed method is tested on the EPANET implementation of the water network of a medium size Danish town (Randers) and shown to perform as intended under varying conditions.
\end{abstract}

%% file: intro.tex
\section{Introduction}
Water distribution networks (WDNs) deliver drinkable water from water sources to consumers using elements such as pumps, pipes, tanks etc. About $7\%-8\%$ of the world's energy is used for water production and distribution \citep{WSSEnergy}. Water pumps account for a significant part of the energy required for water distribution with their percentage ranging from $90\%$ to $95\%$ of the total \citep{ABDELSALAM2021e07820}. There have been many works to schedule the operation of the pumps in WDNs with proper methods so as to reduce energy costs. However, pump scheduling is not an easy task because of the nonlinearities governing the network elements. The problem gets complicated with increasing network size. Also, there are constraints to be satisfied such as limits on tank levels.

In the literature, WDNs with both constant and variable speed  pumps are studied extensively. The control input is turning on and off the pump for constant speed pumps.   In \cite{Chase2009OptimalPS}, a constant-speed pump scheduling problem is posed as an optimization problem in which the decision variables are the operation times of the pumps and the objective is the energy cost.
After observing that the optimal solution would be not running the pumps at all without the constraints, the authors try to find the solution closest to the origin that also complies with the constraints. The proposed way to find such a solution is using a Genetic Algorithm (GA). In \cite{Bagirov2013AnAF}, the Hooke-Jeeves method is used for finding optimal pump operating times for a similar problem. Then, network simulation algorithms are used to check if the constraints are satisfied.  In \cite{CastroGama2017PumpSF}, binary decision variables are used to represent  the opening and the closing of each pump.  The feasibility of the solution found with GA is checked with EPANET, a WDN modeling software, and a high cost is assigned to the infeasible solutions. The number of open pumps is also taken as the input to the system in some works, e.g.,  \cite{Wang2021MinimizingPE}. The problem is then solved using mixed-integer nonlinear programming.  In \cite{Berkel2018AMA}, a network in which pressure zones are connected via constant speed pumps is considered. Each pressure zone is treated as a subsystem and distributed model predictive control (DMPC) is applied. 

The flow rate of the pumps should be determined for networks with variable speed pumps. In \cite{Pour2019EconomicMC}, Linear Parameter Varying (LPV) system modeling is used to replace the nonlinear part of the network, and an Economic Model Predictive Control (EMPC) is applied on top of the LPV system to find the optimal flow rates.  In \cite{Kallese2017PlugandPlayMP}, a network structure with an elevated reservoir is considered. Available data is used for the identification of a reduced system model. Then, EMPC is applied to the model. In the EMPC formulation, node pressures are not constrained. It is assumed that the pressures would be in the accepted range because there is an elevated reservoir. Relaxation of the original problem into a simpler one is commonly used because of the large network sizes. The relaxation is generally achieved by approximating the nonlinear pipe equations with some sort of linear equations or inequalities. In \cite{Baunsgaard2016MPCCO}, pipe equations are linearized around an operating point, and model predictive control (MPC) is applied.  In \cite{Wang2018EconomicMP}, an EMPC is applied to a network where the nonlinear pipe equations are relaxed into a set of linear inequalities. Before simplifying the system model, the network structure is also simplified in \cite{Fiedler2020EconomicNP}. A hierarchical clustering method is used to represent the original network with a smaller one which originally had 378 junctions. A system model is derived from the simplified structure using a Deep Neural Network (DNN) structure. Lagrangian relaxation is used to approximate the original problem in \cite{Ghaddar2015ALD}.  

In this paper, a way for  optimal pump scheduling of large-scale WDNs is presented. To control the pumps, a linear model of the system is derived. Then, a predictive control method with a periodic horizon is constructed. Barrier functions are utilized to prevent constraint violation due to the model-plant mismatch. With the introduction of the periodic horizon and the terminal state constraint, the chance of finding a feasible solution is increased by keeping trajectories close to an optimal periodic trajectory. The method is applied to a medium-sized Danish town's network (Randers).

 The outline of the rest of the paper is as follows. The network model is given in Section \ref{sec:network}. The proposed control method is explained in Section \ref{sec:control}.
 The experimental results are presented in Section \ref{sec:application}. The paper is concluded with final remarks in Section \ref{sec:conc}.

%% file: System.tex
\section{Network Model}
\label{sec:network}
A typical water distribution network consists of pipes, pumps, tanks, junction nodes and reservoirs. Water in the network flows from high hydraulic head to low head. Hydraulic head is a measure of the fluid pressure and is equal to the height of a fluid in a static column at a point. 

Hydraulic head loss occurring in a pipe can be approximated by the Hazen-Williams Equation as
\begin{equation}
\Delta h=h_1-h_2=Kq\lvert q \rvert ^{0.852} 
\label{eq:Haz-Will}
\end{equation}
where $K$ is the pipe resistance that depends on the physical features of a pipe such as diameter and length, $q$ is the flow rate, and $h_1$ and $h_2$ are the heads at the two ends of the pipe.

At each node $j$, the mass conservation law is satisfied. It can be expressed as  
\begin{equation}
 \sum_{i \in \mathpzc{N}_j} q_{ij} = d_j
 \label{eq:NodeMass}
\end{equation}
where $q_{ij}$ is the flow entering the node $j$ from node $i$ and $d_j$ is the demand at node $j$, which is the water requested by the user at node $j$. The symbol $\mathpzc{N}_j$ denotes the set of neighbor nodes of node $j$. Note that  $q_{ij}$ is positive if the flow is from node $i$ to the neighbor node $j$ and negative vice versa. 

Tanks are storage elements that provide water to the users. In the network, tanks are elevated so that water can be pressurized enough to be delivered to the consumers. The change in the water level of a tank is dependent on the flow coming from neighbor nodes and can be written for the tank $j$ as
\begin{equation}
 A_j\dot{{h}}_j = \sum_{i \in \mathpzc{N}_j} q_{ij}
 \label{eq:tankLevel}
\end{equation}
where $A_j$ is the cross-sectional area, $h_j$ is the level of the tank. Tank levels change according to the flow passing through the pipes connected to the tanks. Those flows are determined by a set of pipe head loss equations \eqref{eq:Haz-Will}, and mass balance equations \eqref{eq:NodeMass} throughout the whole network. As Equation \eqref{eq:Haz-Will} is nonlinear, flow through pipes connected to the tanks are nonlinear functions $f_i$ of the demand at each node, tank levels, and the amount of water coming from the pumps. Explicit forms of those nonlinear functions could be derived if the vector $d=[d_1, d_2 ...]^T$ containing the demands of all the nodes is known, which is not possible unless demand data for all nodes are available. In our work, we assume that the total demand of the zones that are supplied by the pumps can be estimated through available data with time series analysis methods, but not require $d$ vector to be known. Since $f_i$ functions can not be found without $d$ vector, we approximate them using linear models and write tank level change equations as 
\begin{equation}
 \dot{{h}}(t)= Ah(t) + B_1u(t) + B_2 d_{a}(t)
 \label{eq:reducedModel}
\end{equation}
where $h(t) \in \mathbb{R}^{n}$ includes tank levels, $A \in \mathbb{R}^{n\times n}$, $B_1 \in \mathbb{R}^{n\times m}$, $B_2 \in \mathbb{R}^{n\times 1}$ are constant system matrices and $d_{a}(t)$ is the aggregated demand of controlled zone at time $t$, $u(t) \in \mathbb{R}^{m}$ is the input containing pump flows. The reason we chose a linear model is to increase the chance of finding a feasible solution for the controller which is posed as an optimization problem in the next section. Although capturing the full dynamics of a large-scale network is not possible with a linear model, the proposed control method is designed to compensate for model inaccuracies and we have observed that it was enough to control the system while satisfying the constraints.

%% file: Control.tex
\section{Periodic Horizon Control}
\label{sec:control}
In this section, a predictive control algorithm for pump scheduling is presented to minimize the economical costs. The aim is to run the pumps when the electricity price is low and let tanks deliver water when the price is high while also satisfying input and output constraints. The problem at time $t$ is posed as 
\begin{subequations}
\label{eq:MPCForm}
\begin{align}
 &\min_{u_0^t,u_1^t \cdots u_{N(t)-1}^t}  \sum_{j=0}^{N(t)-1} J(h_j^t,u_j^t)
\\
 &h_j^t=A_dh_{j-1}^t+B_{d1}u_{j-1}^t+B_{d2} d_{a}(j-1)
 \label{eq:discState}
 \\
 &h_0^t=h(t)
\\
 &u_j^t  \in \mathcal{U} \subseteq \mathbb{R}^{m}&
 \\
 &h_j^t  \in \mathcal{H} \subseteq \mathbb{R}^{n}& \label{eq:MPCstateCons}
 \\
&h_{N(t)}^t \in \mathcal{H}_{tf} \subseteq \mathbb{R}^{n}
\end{align} 
\end{subequations}
where $J(h_j^t,u_j^t)$ is the economic cost function, $h^t=[h_1^t\cdots h_{N(t)}^t ] \in \mathbb{R}^{n\times N(t)}$ is the predicted future states, $u_j^t$ is the input vector, $N(t)$ is the prediction horizon, $\mathcal{U} \subseteq \mathbb{R}^{m}$ and $\mathcal{H} \subseteq \mathbb{R}^{n}$ denotes the input and state constraints respectively and $\mathcal{H}_{tf} \subseteq \mathbb{R}^{n}$ is the terminal state set. The continuous system \eqref{eq:reducedModel} is discretized and \eqref{eq:discState} is obtained. The optimization problem \eqref{eq:MPCForm} is solved at every time step separated by $\Delta_t$ and the first term $u_0^{t}$ of the optimal input sequence $\mathbf{u^t}=[u_0^{t} \cdots u_{N(t)-1}^t] \in \mathbb{R}^{m\times N(t)}$ is applied to the system.  

Input constraints come from the physical limitations and working principles of the pumps. A pump can not provide water in the opposite direction and it can deliver a maximum amount of water per unit of time. These conditions are expressed as 
\begin{equation}
 \mathcal{U}=\{[u_1,\cdots u_m] \in \mathbb{R}^{m} \mid 0\leq u_1 \leq \overline u_1, \cdots 0\leq u_m \leq \overline u_m \}
 \label{eq:inputConstr}
\end{equation}
where $\overline u_1 \cdots \overline u_m$ are upper flow limits. Tank levels are also constrained so that there is always enough water in the tanks in case of an emergency and there is no overflow of water. The set $\mathcal{H}$ can be defined as
\begin{equation}
 \mathcal{H}=\{[h_1,\cdots h_2] \in \mathbb{R}^{n} \mid \tilde h_1\leq h_1 \leq \overline h_1, \cdots \tilde h_n \leq h_n \leq \overline h_n \}
 \label{eq:stateConstr}
\end{equation}

The cost function includes the electricity costs of the pumps. The power provided to the network by the pump  $i$ is equal to ${q}_{pi}(p^{out}_i-p^{in}_i)$, where ${q}_{pi}$ is the pump flow, $p_{out}^i$ and $p_{in}^i$ are the outlet and inlet pressures of the pump $i$. The inlet pressures $p^{in}=[p^{in}_1, p^{in}_2]$ are the pressures of the related reservoirs and are assumed to be constant. The outlet pressures $p^{out}=[p^{out}_1, p^{out}_2]$ are given as the output of the linear model
\begin{equation}
p^{out}(t)=A_ph(t)+B_pu(t)
\label{eq:heads}
\end{equation}
where $A_p$ and $B_p$ are found using system identification on data generated by the EPANET model. Electricity cost at time $t$ is then found by multiplying total power consumption $u(t)^T(p^{out}(t)-p^{in}(t))$ with the electricity price $c(t)$.

We acknowledge a certain degree of model-plant mismatch by using a linear model \eqref{eq:reducedModel} to represent the whole network. This causes actual states $h(t)$ to be different than the predicted states $h^t$. We know that the predicted states satisfy the state constraints \eqref{eq:stateConstr} since they are the solution to the optimization problem \ref{eq:MPCForm}, but the actual states might violate them. To ensure the satisfaction of the state constraints with
the model-plant mismatch, we introduce new terms to the cost function. First, we rewrite state constraints \eqref{eq:stateConstr} as
\begin{equation}
C_i(h) \leq 0, \quad i=0,1,\cdots 2\times n-1
\label{eq:stateIneq}
\end{equation}
where $C_0(h)=\tilde h_1-h_1$ and the rest of the $C_i$ functions are chosen in a similar manner. The cost function terms are then defined as
\begin{equation}
J_{hi}(h)=e^{a_i(C_i(h)+b_i)} \quad i=0,1,\cdots 2\times n-1
\label{eq:barrCost}
\end{equation}
where $a_i,b_i \in \mathbb R_{> 0}$. This can be seen as an exponential barrier function. The parameters $a_i, b_i$  determine a dangerous region close to the boundaries of the state constraints where cost function $J_{hi}$ attains high values. The predicted optimal state trajectories $h^t$ do not enter the dangerous region if possible because of the high cost values in the dangerous region. Then, the actual states $h(t)$ do not violate the state constraints \eqref{eq:stateConstr} assuming the difference between the predicted state and the actual state is small. If the state trajectory enters one of the dangerous regions at any step due to the model-plant mismatch, then the cost function will try to drive the trajectory out of the region.  

The overall cost function includes both the electricity expense term and the constraint barrier functions and it can be expressed as  
\begin{equation}
J(h(t),u(t))=c(t)u(t)^T(p^{out}(t)-p^{in}(t))+\sum_{i=0}^{2\times n-1}J_{hi}(h(t))
\label{eq:CostFunc}
\end{equation}
Both electricity price $c(t)$ and total water demand $d_{a}(t)$ signals can be viewed as consisting of a periodic signal with a period of 1 day and a relatively small deviation signal. This can be leveraged to find a feasible controller. Suppose a sequence of inputs can be found for some initial tank levels such that levels after 1 day are equal to initial levels. In that case, the problem after 1 day is the same as in the beginning assuming deviation signals of the electricity price and the demand are zero, hence they are periodic. Then, the input sequence from the previous day could be applied and produce the same path for tank levels. Taking into account the deviation signals and supposing that a solution exists such that levels after 1 day are close to initial levels, the input sequence from the previous day could be a good point of start to search for a feasible solution if the map from the initial conditions and the demand profile to the optimal input sequences is continuous. Therefore, we choose a terminal state constraint for the end of each day to increase the chance of finding a feasible solution.
\begin{figure}
\resizebox{0.8\columnwidth}{!}{\input{figures/MPCfig.tikz}}
\centering
\caption{Predicted state trajectories $h^t,h^{t+\Delta_t}$ at times $t,t+\Delta_t$. Sampling time $\Delta_t$, prediction horizons $N(t),N(t+\Delta_t)$ and the applied inputs $u_0^t,u_0^{t+\Delta_t}$ are shown. The true state $h(t+\Delta_t)$ and the predicted state $h_1^t$ are indicated to emphasize the deviation from the prediction. The terminal set $\mathcal{B}_r(h_{T_{day}/\Delta_t}^*)$ is also illustrated.}
\label{fig:MPC}
\end{figure}
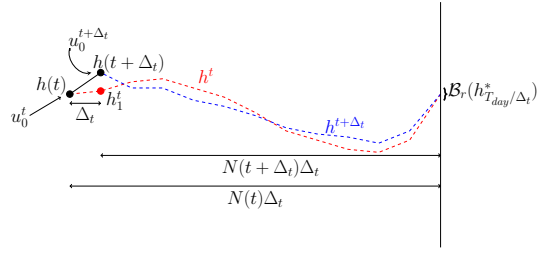
Now, the remaining problem is to decide which tank levels should the trajectories turn back to at the end of each day. We define the optimal periodic trajectory of the system as the solution of
\begin{subequations}
\label{eq:optTraj}
\begin{align}
 &(\mathbf{u^*},\mathbf{h^*})=\argmin_{{u_i},{h_i}}  \sum_{i=0}^{(T_{day}/\Delta_t)-1} J({h_i},{u_i})
\\
 &{h_i}=A_d{h_{i-1}}+B_{d1}{u_{i-1}}+B_{d2} d_{a}^*(i-1)
\\
 &{u_i}  \in \mathcal{U} \subseteq \mathbb{R}^{m}&
 \\
 &{h_i}  \in \mathcal{H} \subseteq \mathbb{R}^{n}&
 \\
 &{h_0}=h_{T_{day}/\Delta_t}
  \label{eq:period}
\end{align}
\end{subequations}
where $T_{day}$ is the duration of a whole day, $d_{a}^*$ is the average daily demand profile obtained from the past measurements. The resulting state trajectory $\mathbf{h^*}=[h^*_0 \cdots h^*_{T_{day}/\Delta_t}] \in \mathbb{R}^{n\times(T_{day}/\Delta_t+1)}$ is the optimal periodic trajectory  because of the constraint \eqref{eq:period}. The terminal set $\mathcal{H}_{tf}$ and the prediction horizon $N(t)$ is chosen to make tank levels at the end of each day close to $h_{T_{day}/\Delta_t}^*$. At any time $t$, $t+N(t)\Delta_t$ should be equal to the end of the day. $\mathcal{H}_{tf}$ and $N(t)$ could be written as
\begin{subequations}
\label{eq:terminalandHorizon}
\begin{align}
\label{eq:terminal}
 &\mathcal{{H}}_{tf}= \mathcal{B}_r( h_{T_{day}/\Delta_t}^*)
 \\
 &N(t)=(T_{day}-t\bmod T_{day})/\Delta_t
\end{align}
\end{subequations}
where $\mathcal{B}_r(h_{T_{day}/\Delta_t}^*)$ is the open ball centered at $h_{T_{day}/\Delta_t}^*$ with radius $r$.
 Note that $N(t)$ changes so that $t+N(t)\Delta_t$ is the end of the day for all $t$.  With these definitions, the condition \eqref{eq:terminal} will translate to tank levels at the end of the day being close to the final point in optimal periodic trajectory $h_{T_{day}/\Delta_t}^*$ as shown in Figure \ref{fig:MPC}. Therefore, not only chance of finding a feasible solution is increased but also the solutions are kept around the optimal periodic trajectory $\mathbf{h^*}$. If the problem \eqref{eq:MPCForm} becomes infeasible at any time step $t$, we apply the second term of the input sequence from the previous step $u_1^{t-\Delta_t}$. The reason behind this choice is as follows: If we apply the optimal control input $u_0^{t-\Delta_t}$ to the network model \eqref{eq:reducedModel} at time $t-\Delta_t$, then the optimal sequence in the next time step will be $\mathbf{u^t}=[u_1^{t-\Delta_t} \cdots u_{N(t-\Delta_t)-1}^{t-\Delta_t}]$ following Bellman's principle of optimality. Then, at time $t$,  $u_1^{t-\Delta_t}$ will be applied to the system as calculated at $t-\Delta_t$. Assuming the model-plant mismatch is small enough, $u_1^{t-\Delta_t}$ is still a good input candidate if the problem is infeasible at time $t$.

%% file: figures/MPCfig.tikz
\begin{tikzpicture}
    \begin{scope}[yscale=-1,xscale=1]
        \draw[rotate=-270,dashed,red] (0,0) --  (1.5,1) --
           (1.9,2) --  
          (1.8,3)  -- (1.5,4) --
           (1.2,5) -- 
          (0.6,6)  -- (0.1,7) --
           (-0.2,8) -- 
          (-0.5,9) --  (-0.4,10) --
           (-0.1,11) -- 
          (0,12);

          \draw[rotate=-270,dashed,blue] (0,0)  -- (1.2,1) --
           (1.6,2) -- 
          (1.4,3) --  (1.3,4) --
           (1.1,5) -- 
          (0.7,6) --  (0.4,7) --
           (0.2,8) -- 
          (-0.2,9) --  (-0.2,10) --
           (-0.7,11)  ;
         \draw[rotate=-270] (0,12)-- (-0.7,11);
         
    \end{scope}

    \node (p1) at (-12,0){};
     \filldraw[black] ($(p1)$) circle (3pt)      node[anchor=west]{};
     \node (p2) at (-11,0.1){};
     \filldraw[red] ($(p2)$) circle (3pt)      node[anchor=west]{};
      \node (p3) at (-11,0.7){};
     \filldraw[black] ($(p3)$) circle (3pt)      node[anchor=west]{};
     \draw (0,3) -- (0,-5);
     
     \draw [<->] (-12,-0.3) -- (-11,-0.3)
                                node [below,midway,draw=none] {\Large$\Delta_t$};
    \draw [<->] (-12,-3) -- (0,-3)
                                node [below,midway,draw=none] {\Large $N(t)\Delta_t$};
     \draw [<->] (-11,-2) -- (0,-2)
                                node [below,midway,draw=none] {\Large $N(t+\Delta_t)\Delta_t$};                           
    \node[draw=none,align=left] at (-12.5,0.3) {\Large$h(t)$ };
    \node[draw=none,align=left] at (-10,1) {\Large$h(t+\Delta_t)$ };
    \node[draw=none,align=left] at (-10.45,-0.2) {\Large$h_1^t$ };

     \draw [->]   (-13.3,-0.7) -- (-12.3,-0.1)
                                node [midway, right, xshift=-35pt, yshift=-10pt,draw=none] {\Large$u_0^{t}$};
      \draw [brace mirrored, thick] (-12,1.5) edge[bend right=60,-stealth]    (-11.3,0.7)
                                node [ right, xshift=-5pt, yshift=8pt,draw=none] {\Large$u_0^{t+\Delta_t}$};         
      \node[draw=none,align=left,blue] at (-3,-1) {\Large$h^{t+\Delta_t}$ };
      \node[draw=none,align=left,red] at (-7.5,0.6) {\Large$h^{t}$ };
     \draw[brace mirrored, ultra thick] (0.1,-0.2)--(0.1,0.2) node [midway, right] {\Large$\mathcal{B}_r(h_{T_{day}/\Delta_t}^*)$};
\end{tikzpicture}

%% file: exp.tex
\section{Application}
\label{sec:application}
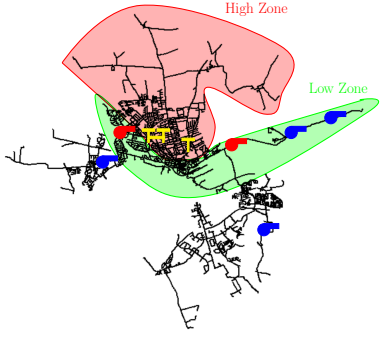
\begin{figure}
\resizebox{0.6\columnwidth}{!}{\input{figures/networkIm.tikz}}
\centering
\caption{Water Distribution Network of Randers. The pumping stations to be controlled are shown in red. Tanks are shown with a 'T' shaped symbol in yellow.}
\label{fig:network}
\end{figure}
The presented method is applied to WDN of Randers, a Danish city, which is shown in Figure \ref{fig:network}.  The network consists of 4549 nodes and 4905 links connecting them. There are 8 pumping stations in the network, 6 of which are shown in the figure whereas the other 2 are stationed where tanks are placed. The goal is to derive the schedules for 2 of the pumping stations while other pumps are already working according to some predetermined strategies. The stations to be controlled are shown in red in the figure. Their task is to deliver water mostly to the High Zone (HZ) and Low Zone (LZ). However, connections exist between HZ-LZ and the rest of the city, so we can not think of the system as composed of isolated networks entirely. There are also 3 tanks in the HZ. While 2 of them are directly connected via pipes, the third one stands alone as shown in the figure.  

\begin{figure}
\includegraphics[width=7.5cm]{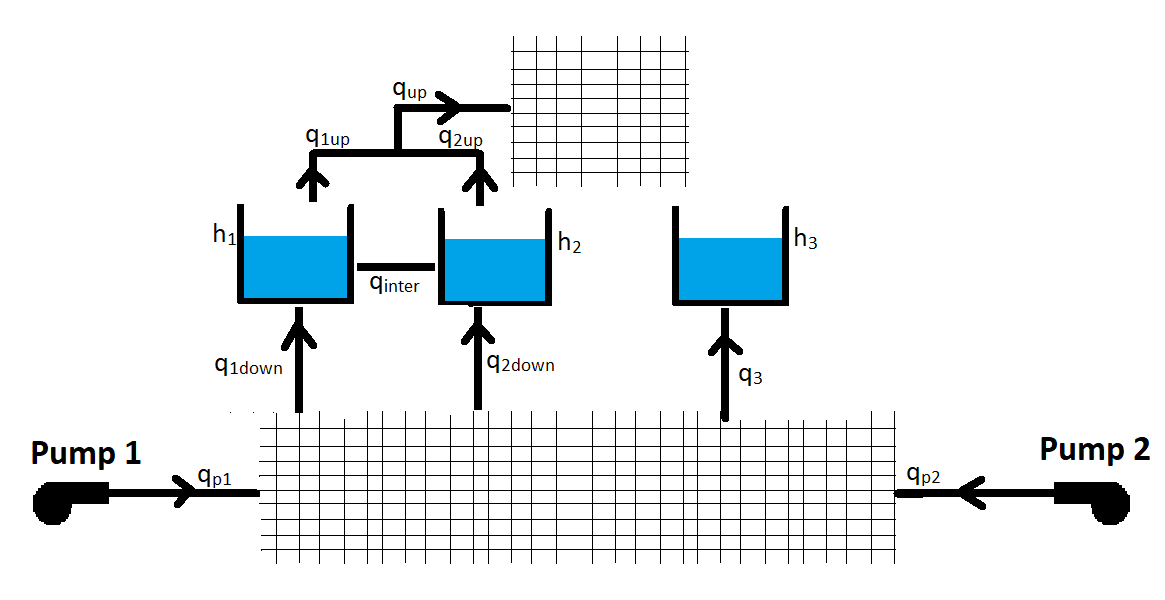}
\centering
\caption{Structure of the WDN.}
\label{fig:netModel}
\end{figure}
The overall structure of the Randers WDN with tanks and pumps to be controlled are given in Figure \ref{fig:netModel}. There are 3 water tanks in the network, 2 of which have been connected with a pipe directly.  The tank level changes can be written by applying the mass conservation law \eqref{eq:tankLevel} to the tanks in Figure \ref{fig:netModel}  as
\begin{subequations}
\label{eq:tankLevelEq}
\begin{align}
        \label{eq:1}
         A_1\dot{{h}}_1&= q_{1down}+q_{1up}+q_{inter}\\
            \nonumber&=f_1(h_1,h_2,h_3,q_{p1},q_{p2},d),\\
            \label{eq:2}
         A_2\dot{{h}}_2&= q_{2down}+q_{2up}-q_{inter}\\
        \nonumber&=f_2(h_1,h_2,h_3,q_{p1},q_{p2},d),\\
  A_3\dot{{h}}_3 &=q_{3}=f_3(h_1,h_2,h_3,q_{p1},q_{p2},d),
\end{align}
\end{subequations}
where $d$ is the vector containing the demands of all the nodes, $q_{p1},q_{p2}$ are the pump flows, $A_1,A_2,A_3$ are the cross sectional areas of the tanks and $f_1,f_2,f_3$ are nonlinear flow functions. Water levels at the two connected tanks are almost equal $h_1\thickapprox h_2$ all the time since the pipe connecting respective tanks is big enough to oppose the water flows coming from neighbor nodes. That enables us to consider $h_1,h_2$ together as
\begin{equation}
 (A_1+A_2)\dot{{h}}_{1,2} \thickapprox q_{1down}+q_{2down}+q_{up}=f_1+f_2.
 \label{eq:h1h2connect}
\end{equation}
We have used the EPANET model of the network to generate the data required for approximating $f_1+f_2$ and $ f_3$. The model is simulated with various tank level initial conditions and flow rates of 2 pumping stations to be controlled. The control laws for the remaining pumping stations are already defined in the EPANET model. Then, the linear model \eqref{eq:reducedModel} is fitted to simulation data using least squares. The state variables for the model are $h(t)=[{h}_{1,2}(t),{h}_{3}(t)] \in \mathbb{R}^{2}$ and the inputs are $u(t)=[{q}_{p1}(t), {q}_{p2}(t)] \in \mathbb{R}^{2}$. The total demand of High and Low Zone is used as aggregated demand $d_{a}$ in the model since mainly those areas are supplied by the controlled pumps.

\subsection{Simulation Results}

The proposed control method is tested on EPANET model of Randers water network. Epanet-Matlab toolkit \cite{Eliades2016} is used to set the flow of the 2 pumps at each time step and simulate the network. The remaining pumps are controlled with rule-based control laws that are previously defined on EPANET. 

The parameters of exponential barrier functions $J_{hi}$ are chosen as $a_i=80,$ $b_i=0.3$ for all $i$. It is assumed that the electricity prices are known in advance during the test. Tank levels $h_1,h_2$ have a maximum value of 3m while $h_3$ has 2.8m. Tanks are required to be at least half full. Maximum pump flows are set to 100. Sampling time $\Delta_t$ is set to 1 hour in the experiments, so the control input is recalculated at each hour. We assume that total demand $d_{a}(t)$ of HZ and LZ  can be estimated up to 1 day from available data. Although we do not have historical data on the demand, we imitate this behaviour by using a slightly perturbed version of the real demand used in EPANET simulation during MPC calculations. The perturbations are adapted from a real demand data set of a small Danish facility. Normalized difference between the average demand and the demand of a random day in data set is added to EPANET demand to replicate estimated demand. In each experiment a different day from the data set is used, so the assumed estimated demand is different each time.

\begin{figure}
\centering
\subfloat[]{\includegraphics[width=0.40\textwidth]{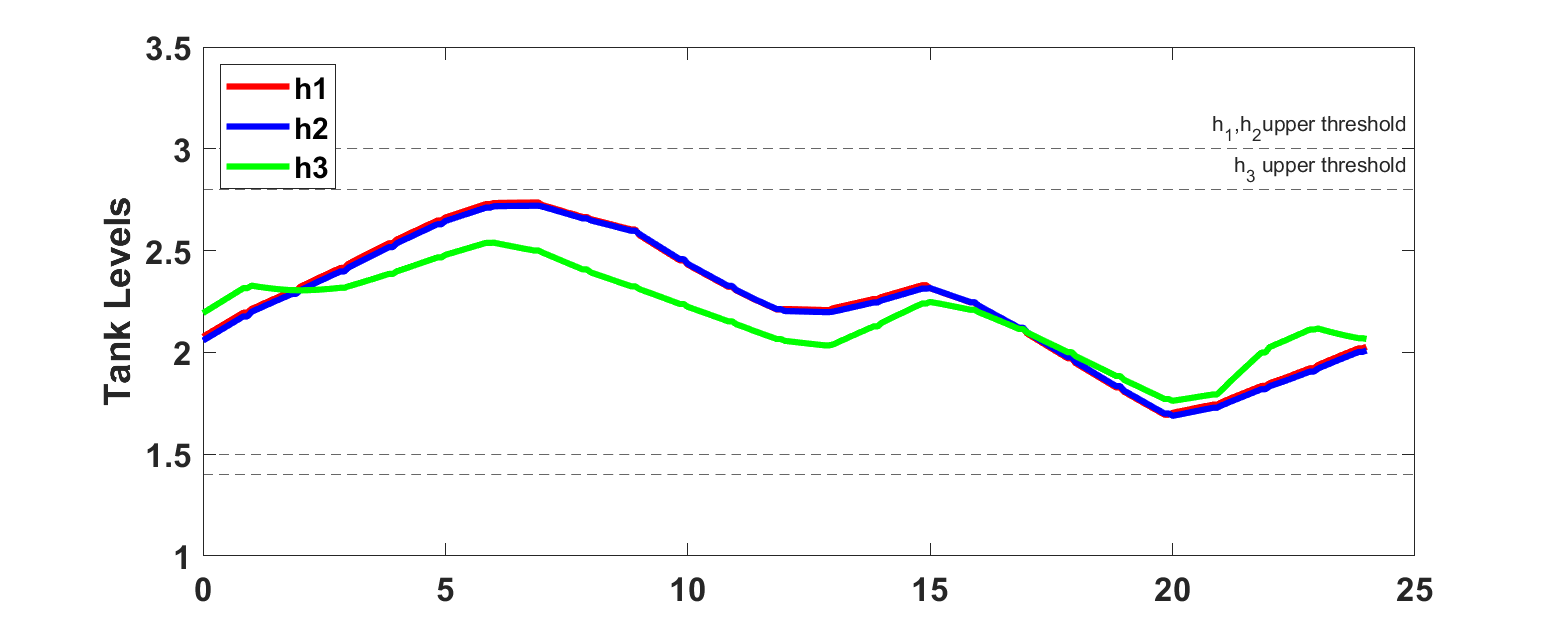}} \\
\subfloat[]{\includegraphics[width=0.40\textwidth]{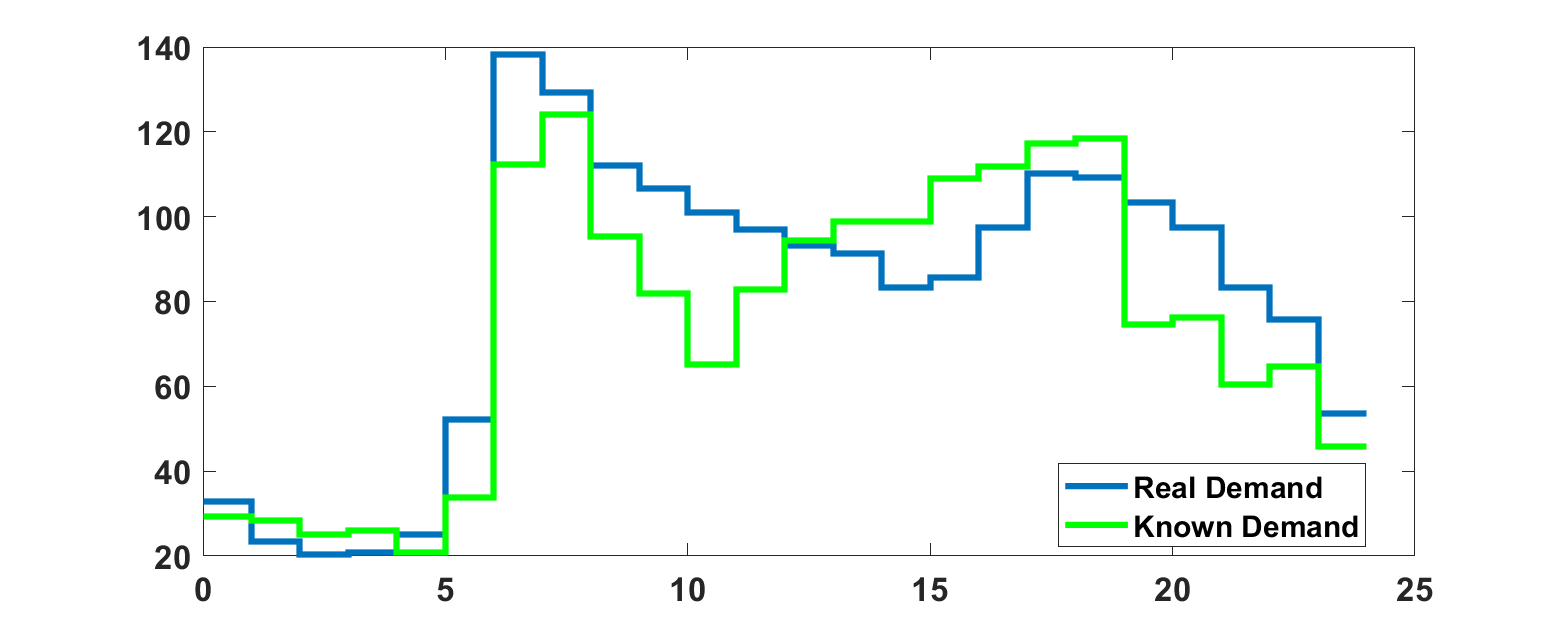}} \\
\subfloat[]{\includegraphics[width=0.40\textwidth]{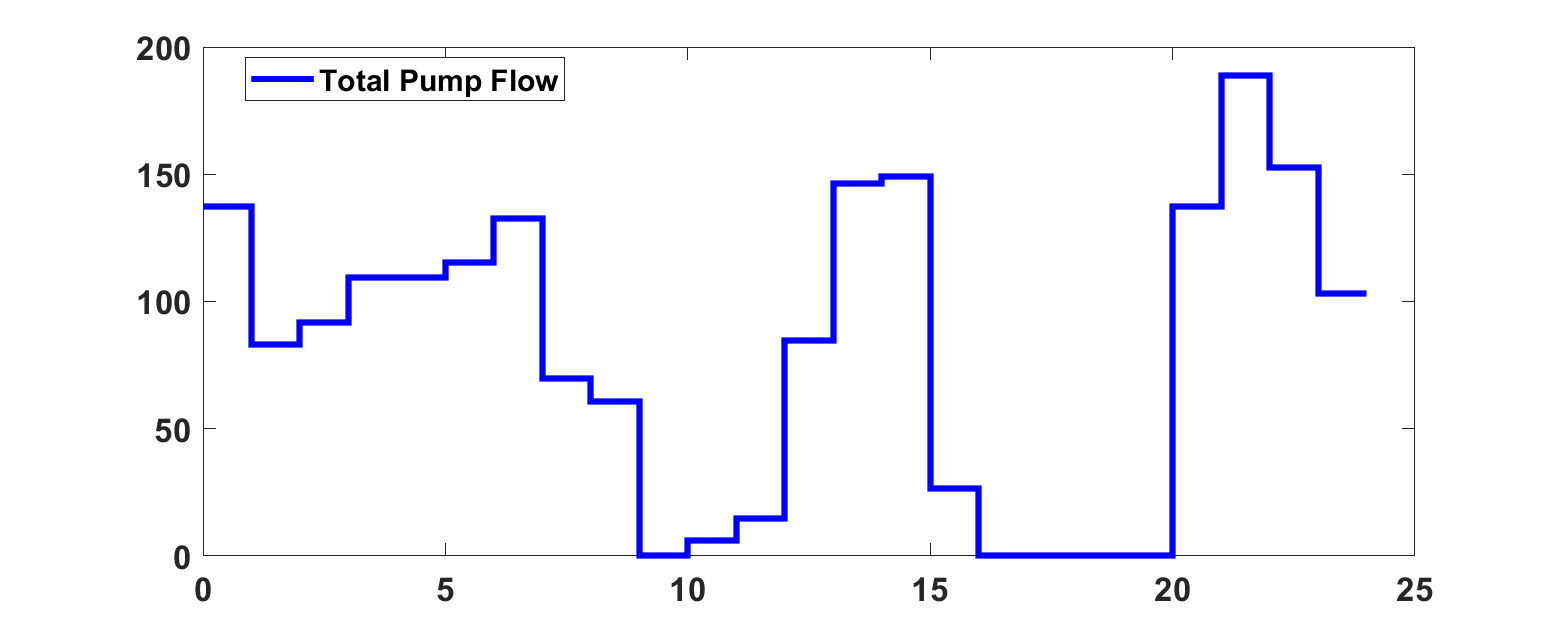}} \\
\subfloat[]{\includegraphics[width=0.40\textwidth]{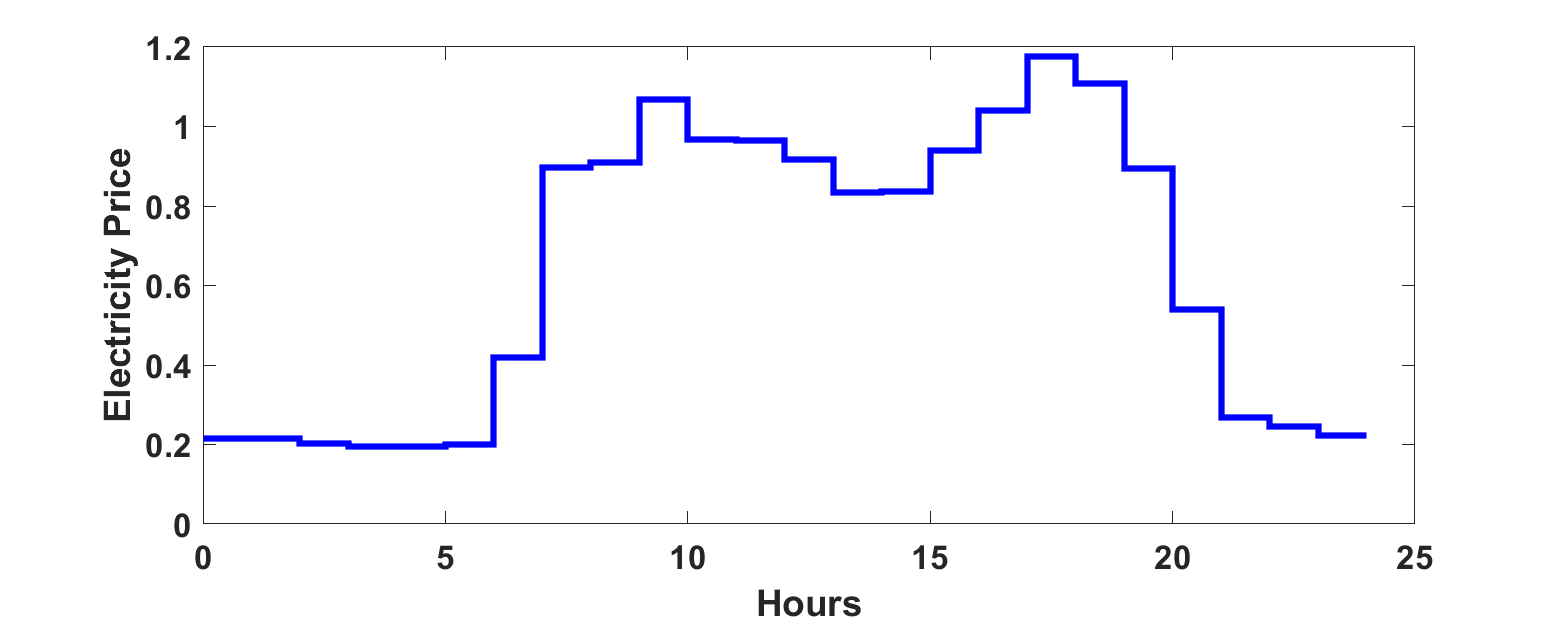}}

\caption{Sample simulation. (a) evolution of tank levels through 1 day with upper and lower level thresholds; (b) real total demand of HZ and LZ used in EPANET simulation and the demand used in MPC calculations; (c) total flow provided by the 2 pumps; (d) electricity price.}
\label{fig:mainresult}
\end{figure}

\begin{figure}
\centering
\subfloat{\includegraphics[width=0.24\textwidth]{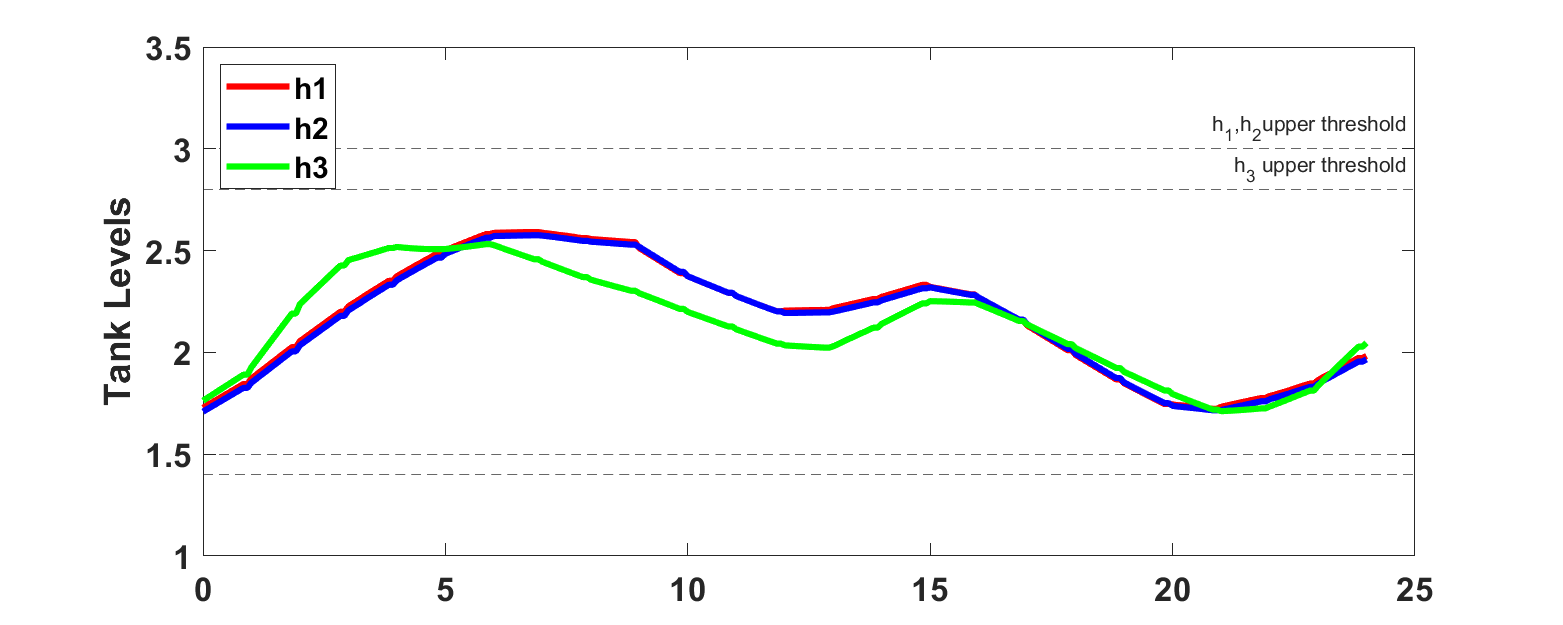}}
\subfloat{\includegraphics[width=0.24\textwidth]{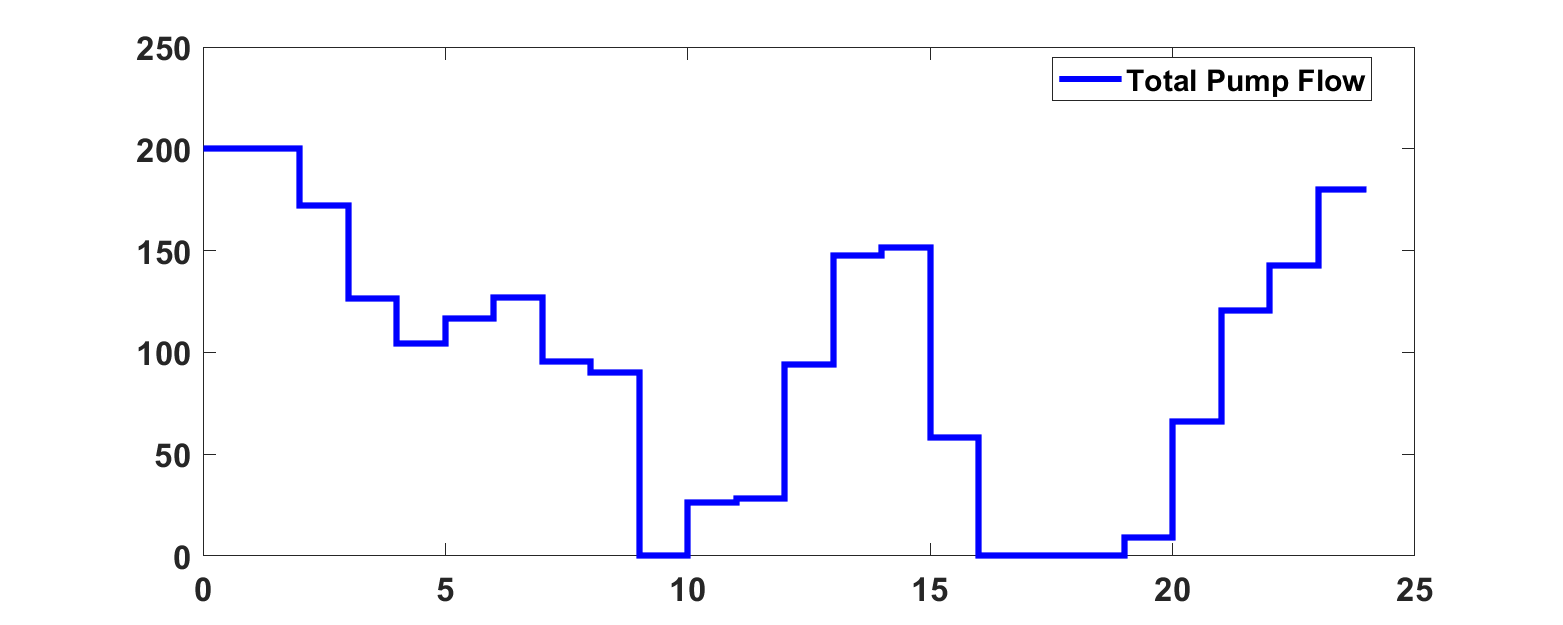}} \\
\subfloat{\includegraphics[width=0.24\textwidth]{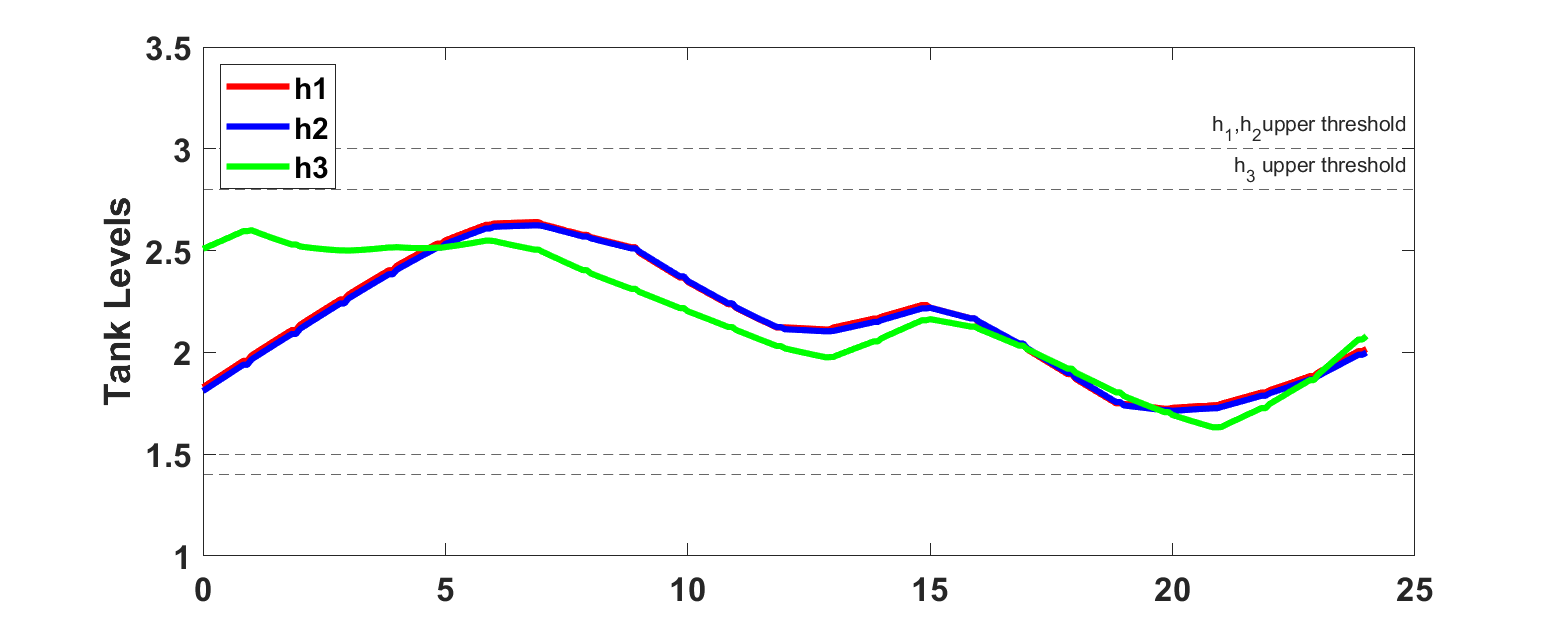}}
\subfloat{\includegraphics[width=0.24\textwidth]{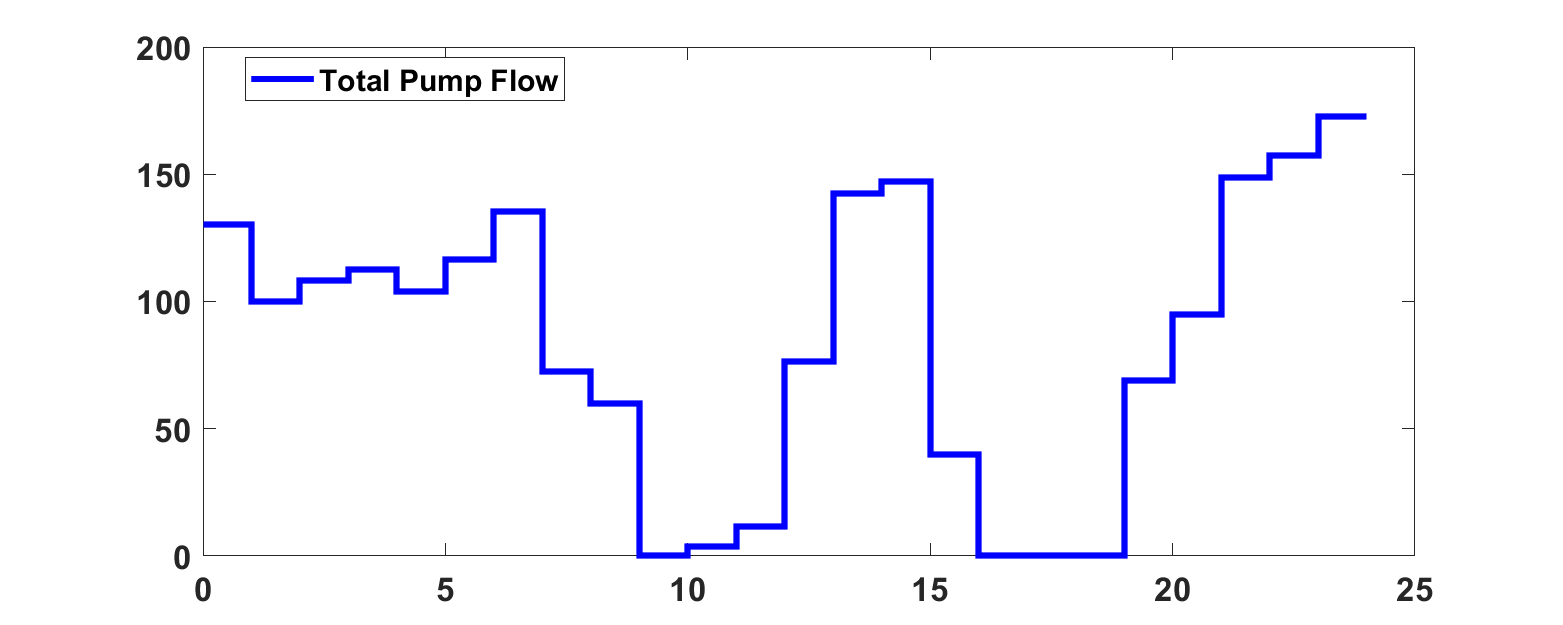}} \\
\subfloat{\includegraphics[width=0.24\textwidth]{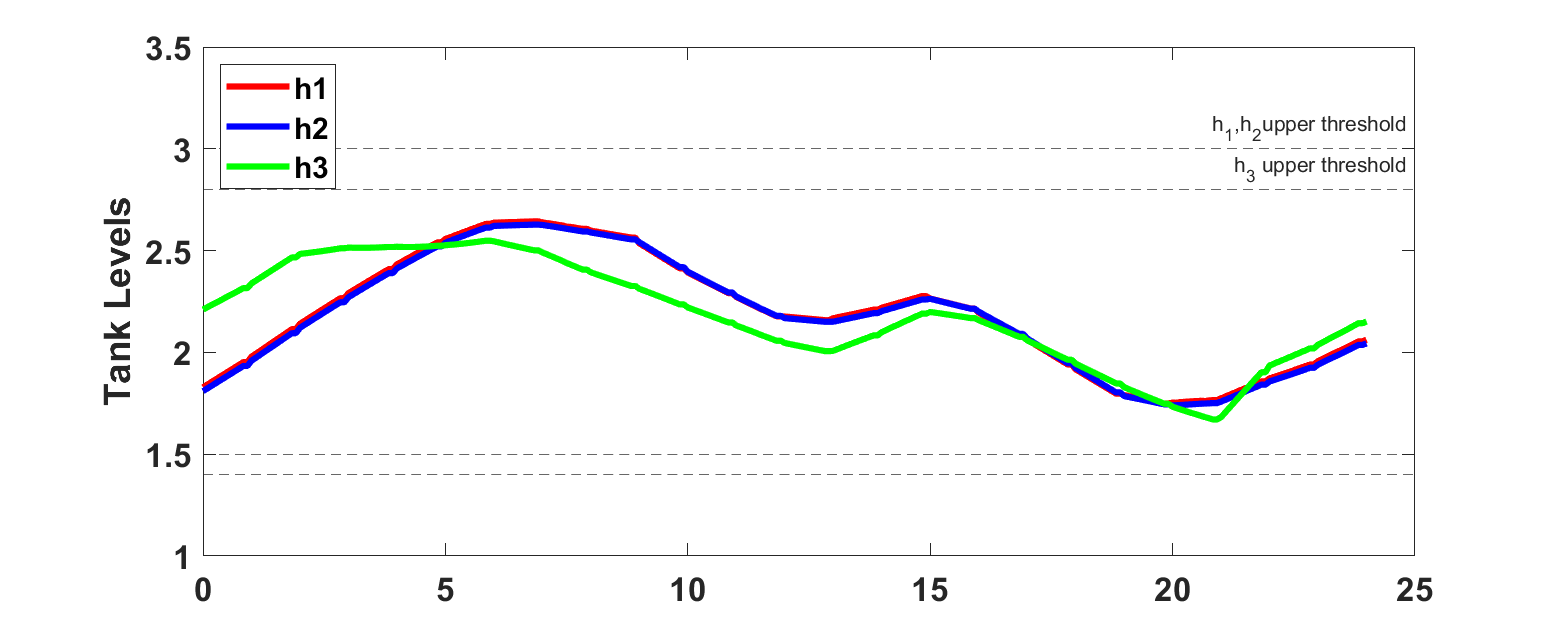}}
\subfloat{\includegraphics[width=0.24\textwidth]{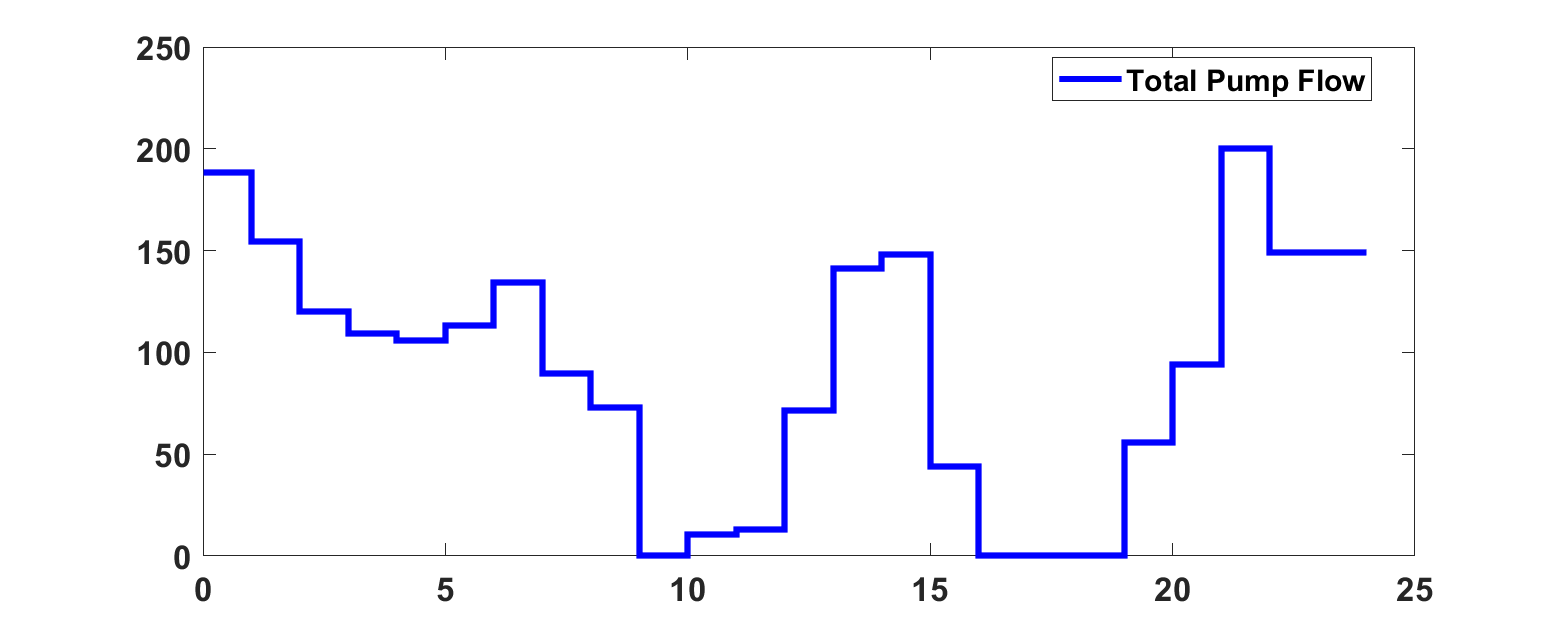}} \\
\subfloat{\includegraphics[width=0.24\textwidth]{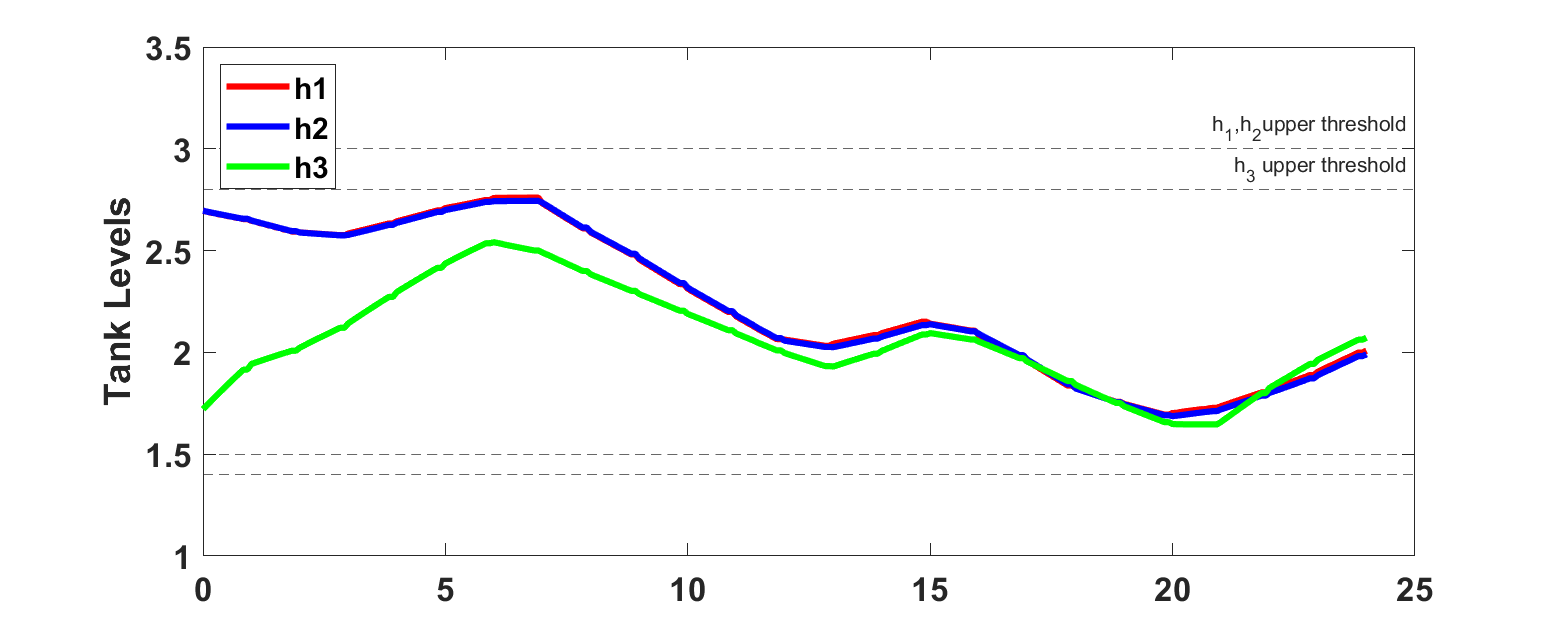}}
\subfloat{\includegraphics[width=0.24\textwidth]{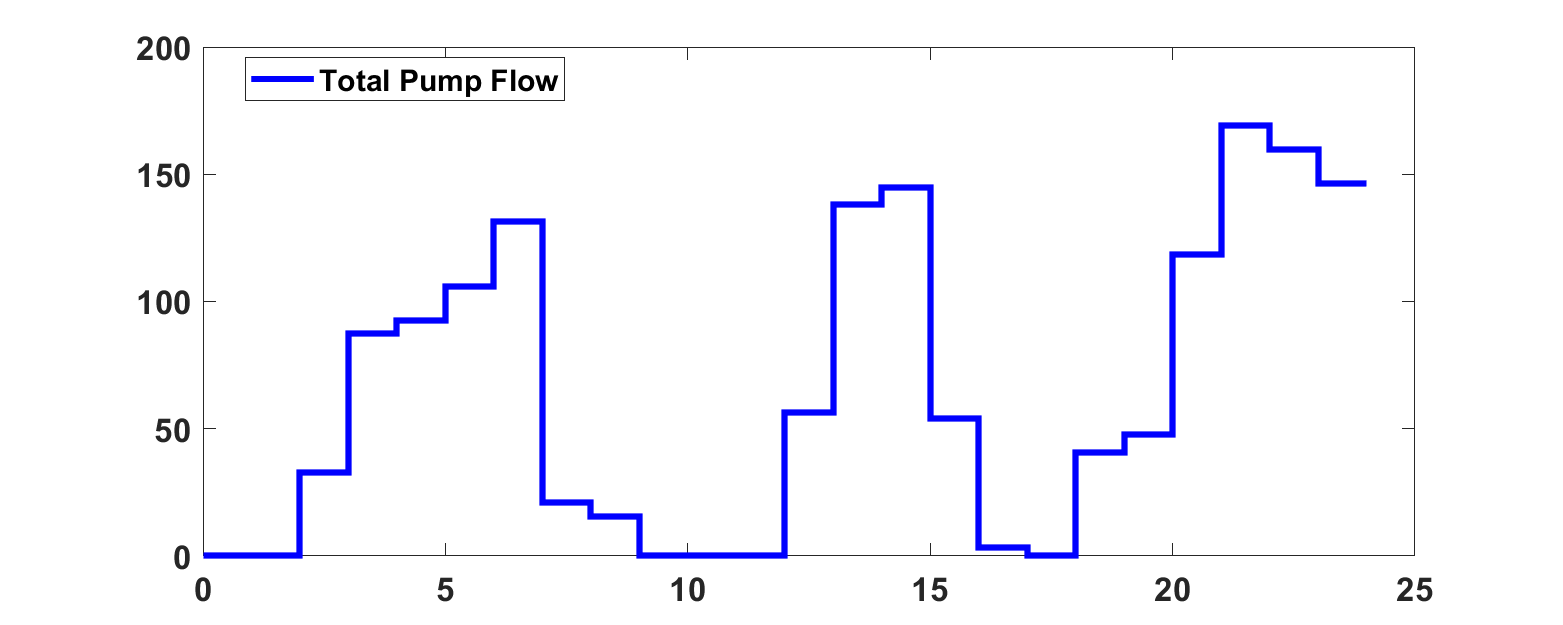}}
    
\caption{Tank levels and pump flows for different initial conditions}
\label{fig:multiplecases}
\end{figure}

The simulation results when the presented method is applied to the EPANET model are given in Figure \ref{fig:mainresult}. The initial tank levels are equal to $h_{T_{day}/\Delta_t}^*$ in the simulation. The top plot shows the evolution of tank levels along with the upper and lower thresholds. It is seen that the thresholds are not violated and moreover tank levels are not getting too close to them, which was the idea behind exponential barrier functions. Both the real demand and the assumed estimated demand of HZ and LZ are in the figure below. Total applied pump flows and electricity prices are in the following figures. The expected result is pump flows being higher when electricity prices are low, and lower when they are high, which seems to be the case as can be seen in the plot. Pump flows drop significantly when prices are at the peak and they reach their highest value at the end of the day when prices are low. A more aggressive controller can be obtained by picking a smaller $b_i$ value for barrier functions at the expense of risking constraint violation. 
In Figure \ref{fig:multiplecases}, the tank level simulation results and control inputs for different initial conditions and different assumed estimated demands are given. The electricity price profile is the same as before. It is seen that the algorithm is able to control the network on various cases while satisfying the constraints. Regardless of initial tank levels, the pumping profiles have a similar profile: high pump flows close to midnight and in the middle of the day. The only exception is the bottom plot. In the beginning, prices are low but pump flows are not high. This can be attributed to water levels $h_1,h_2$ being close to the upper thresholds and water demand being low in the beginning.

\begin{figure}
\includegraphics[width=7.3cm]{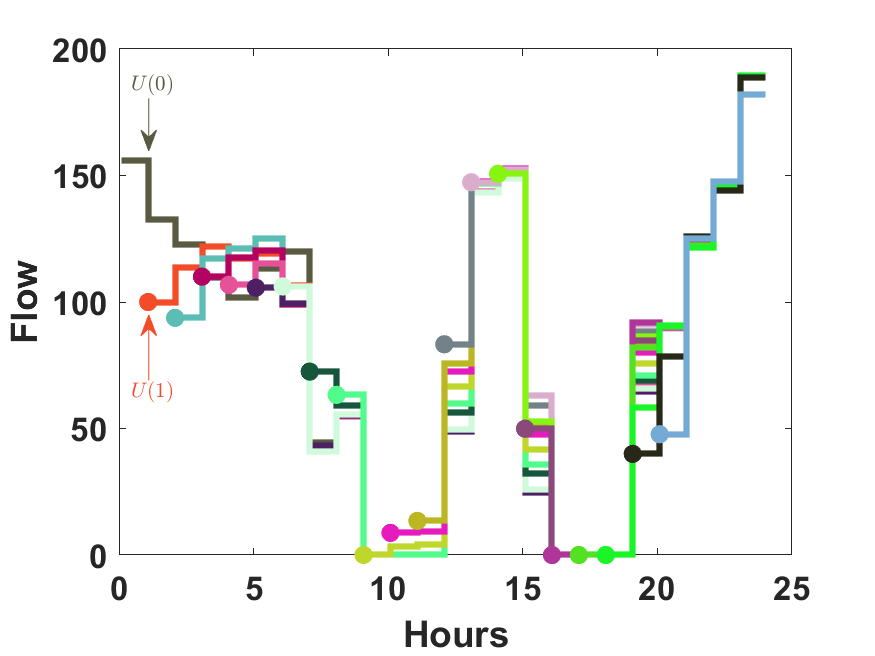}
\centering
\caption{Evolution of found input sequences $U(t)$ through 1 day. It can be seen that the solutions remain close to the initial optimal sequence $U(0)$.}
\label{fig:solutions}
\end{figure}

The assumption that the optimal input sequences $U(t)$ would not diverge a lot from the one found in previous step $U(t-\Delta_t)$ is the reason we apply $u_1^{t-\Delta_t}$ at time $t$ if the problem \eqref{eq:MPCForm} is infeasible at time $t$. This assumption is tested with initial conditions $h_{1,2,3}=h_{T_{day}/\Delta_t}^*$. In figure \ref{fig:solutions}, total pump flow $[1, 1]^T u_i^t, i=0 \cdots N(t)-1$ of the found optimal input sequences $U(t), t=0,\Delta_t \cdots T_{day}-\Delta_t$, except when the problem were infeasible, are given.  It can be seen that $u_1^{t-\Delta_t}$ is close to the $u_0^{t}$ for all $t$, which shows that our assumption is valid at least for this experiment.

Finally, the ability of the algorithm to decrease economic costs is tested with various initial conditions. For each case, a demand follower pumping strategy is used as a benchmark. The flow of the 2 pumps is adjusted with trial and error for each demand follower such that the total flow of the 2 pumps is equal to water demand at each time step and tank levels satisfy the terminal constraint \eqref{eq:terminal}. The demand follower is a natural candidate to be a benchmark method  since providing as much water as demand is an intuitive idea and the constraints in \eqref{eq:MPCForm} can be satisfied with manual adjustments of pump flows. The economic costs are presented relatively in Table \ref{tab:relcost} As it is seen, the proposed algorithm saves between $40\%$ and $45\%$ of the cost with different demand profiles.
\begin{table}
\centering
\begin{tabular}{ c|c } 
  \textbf{Proposed Method} & \textbf{Demand Follower} \\
 \hline
  0.5967 & 1 \\
 \hline
 0.5745 & 1  \\
 \hline
  0.5826 & 1 \\
 \hline
  0.5558 & 1  \\
\hline
\end{tabular}
\caption{Relative economic costs of the proposed method and demand follower strategy for various demand profiles}
\label{tab:relcost}
\end{table}

%% file: figures/networkIm.tikz
\begin{tikzpicture}
    \begin{scope}[transparency group]
        \begin{scope}[blend mode=multiply]
            \node[inner sep=0pt] (russell) at (0,0)
            {\includegraphics[width=15cm]{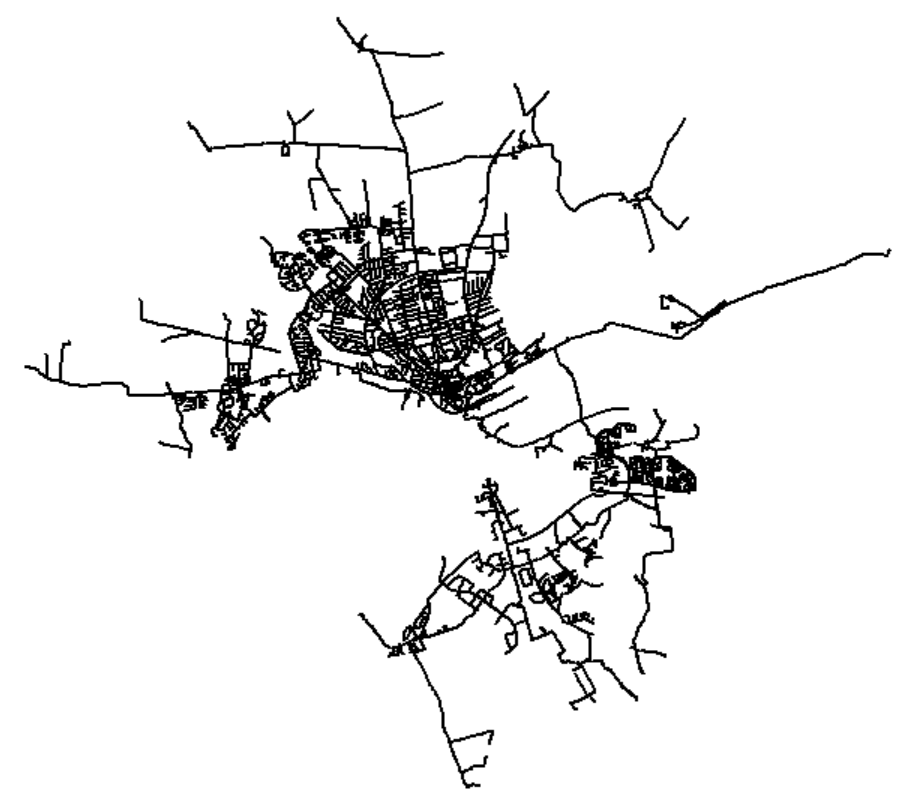}};
            \draw[color=red!200, fill=red!30,  very thick]  plot[smooth, tension=.7] coordinates {(-4.83,4.33) (-1.9159,1.666) (-0.8333, 0.4998)   (-0.1666, 0.4165) (0.7497, 0.4998) (1.3328, 1.666) (0.9996, 3.332) (3.332, 2.2491) (3.9984, 2.9988) (3.9984, 4.998) (-1.666, 6.664) (-4.8314,4.3316)};
            \draw[color=green!200, fill=green!30,  very thick]  plot[smooth, tension=.7] coordinates {(7.3304, 2.8322) (2.3324, 1.3328) (1.1662, 0.833)  (0.9996, 0.4998) (-0.833, 0.4998) (-1.9159,1.666) (-3.4986, 2.9988) (-2.1658, 0) (1.3328, -0.9996) (7.497, 2.3324) (7.1638, 2.8322)  };
        \end{scope}
    \end{scope}
     \node (p1) at (6,2.1){};
     \filldraw[blue] ($(p1)$) circle (7pt) node[anchor=west]{};
    \filldraw[blue, very thick] ($(p1)+(0,0.05)$) rectangle ($(p1)+(0.6,0.25)$);

    \node (p2) at (3.3,-2.45){};
     \filldraw[blue] ($(p2)$) circle (7pt) node[anchor=west]{};
    \filldraw[blue, very thick] ($(p2)+(0,0.05)$) rectangle ($(p2)+(0.6,0.25)$);

    \node (p3) at (4.4,1.5){};
     \filldraw[blue] ($(p3)$) circle (7pt) node[anchor=west]{};
    \filldraw[blue, very thick] ($(p3)+(0,0.05)$) rectangle ($(p3)+(0.6,0.25)$);

    \node (p4) at (-3.2,0.3){};
     \filldraw[blue] ($(p4)$) circle (7pt) node[anchor=west]{};
    \filldraw[blue, very thick] ($(p4)+(0,0.05)$) rectangle ($(p4)+(0.6,0.25)$);

    \node (p5) at (-2.5,1.5){};
     \filldraw[red] ($(p5)$) circle (7pt) node[anchor=west]{};
    \filldraw[red, very thick] ($(p5)+(0,0.05)$) rectangle ($(p5)+(0.6,0.25)$);

    \node (p6) at (2,1){};
     \filldraw[red] ($(p6)$) circle (7pt) node[anchor=west]{};
    \filldraw[red, very thick] ($(p6)+(0,0.05)$) rectangle ($(p6)+(0.6,0.25)$);

    \node (t1) at (-1.6,1.5){};
     \filldraw[yellow, very thick] ($(t1)+(0,0.05)$) rectangle ($(t1)+(0.5,0.15)$);
    \filldraw[yellow, very thick] ($(t1)+(0.22,0.05)$) rectangle ($(t1)+(0.28,-0.4)$);

    \node (t2) at (-1,1.5){};
     \filldraw[yellow, very thick] ($(t2)+(0,0.05)$) rectangle ($(t2)+(0.5,0.15)$);
    \filldraw[yellow, very thick] ($(t2)+(0.22,0.05)$) rectangle ($(t2)+(0.28,-0.4)$);

    \filldraw[yellow, very thick] ($(t2)+(-0.3,-0.15)$) rectangle ($(t2)+(0.3,-0.25)$);

    \node (t3) at (0,1.12){};
     \filldraw[yellow, very thick] ($(t3)+(0,0.05)$) rectangle ($(t3)+(0.5,0.15)$);
    \filldraw[yellow, very thick] ($(t3)+(0.22,0.05)$) rectangle ($(t3)+(0.28,-0.4)$);
    
    \node[text=red, ] at (3,6.5) {\LARGE High Zone};
    \node[text=green, ] at (6.3,3.3) {\LARGE Low Zone};
\end{tikzpicture}

%% file: conc.tex
\section{Conclusion}
\label{sec:conc}
We have presented a predictive control algorithm with a periodic horizon for WDNs. The aim is to minimize the economic cost and satisfy the operational constraints. A linear model is used to represent Randers WDN to increase the chance of finding a solution to the problem \eqref{eq:MPCForm} at expense of a model-plant mismatch. Periodic horizon is introduced to the predictive control formulation to keep the resulting state trajectories around the optimal periodic trajectory. Barrier functions are used to prevent constraint violation since there is a model-plant mismatch.

The presented algorithm is tested on Randers WDN using EPANET. It is shown in various situations that the method is able to find an economic solution where pump flows are adjusted according to electricity prices. Also, it is shown that the system trajectories do not enter dangerous zones introduced by barrier functions as long as the predicted demand and the actual demand are somewhat close.

As future work, we plan to work on theoretical guarantees of the existence of solutions to the proposed method. Also, the robustness of periodic horizon control of periodical systems with barrier functions will be investigated.